\documentclass{IEEEtran}
\usepackage{cite}
\usepackage{amsmath,amssymb,amsfonts}
\usepackage{algorithmic}
\usepackage{graphicx}
\usepackage{textcomp}
\usepackage{tikz}
\usepackage{graphicx} 
\usepackage{layout}
\usepackage{array}
\usepackage{amsmath}
\usepackage{amssymb}
\usepackage{color}
\usepackage{epstopdf}
\usepackage{algorithm} 
\usepackage{algorithmic}
\usepackage{import}
\usepackage{tikz}
\usepackage{pgfplots}
\usepackage{tikz,tikz-3dplot}
\usepackage{bm}
\usepackage{subfig}
\usepackage{array}
\newcolumntype{P}[1]{>{\centering\arraybackslash}p{#1}}
\newcolumntype{M}[1]{>{\centering\arraybackslash}m{#1}}
\usepackage{booktabs}
\usepackage{array}
\newcolumntype{C}{>{\centering\arraybackslash}m{1.4cm}}

\usepackage{tabularx, booktabs, makecell, caption}
    \usepackage{siunitx}

\usepackage{accents}

\usepackage{amsmath}

\makeatletter
\newcommand{\thickhline}{%
    \noalign {\ifnum 0=`}\fi \hrule height 1.5pt
    \futurelet \reserved@a \@xhline
}
\newcolumntype{"}{@{\hskip\tabcolsep\vrule width 1pt\hskip\tabcolsep}}
\makeatother

\usetikzlibrary{arrows.meta}
\usetikzlibrary{calc}
\tikzset{axis line style/.style={thin, gray, -stealth}}

\def\BibTeX{{\rm B\kern-.05em{\sc i\kern-.025em b}\kern-.08em
    T\kern-.1667em\lower.7ex\hbox{E}\kern-.125emX}}
\begin{document}
\title{A Fast Direct Solver for Mutual Coupling Analysis of Large Arrays of Reflector Antennas}
\author{Quentin~Gueuning, Eloy de Lera Acedo, Anthony Keith Brown, \emph{Life Senior Member, IEEE}, and Nicolas Fagnoni 

\thanks{Q. Gueuning and E. de Lera Acedo are with the Astrophysics Group, Cavendish
Laboratory, University of Cambridge, UK and also with the Kavli Institute for Cosmology, Cavendish
Laboratory, University of Cambridge, UK.}
\thanks{A. K. Brown is with the School of Electronic Engineering and
Computer Science, Queen Mary University of London, London, UK
and also the University of Manchester, UK.}}

\markboth{IEEE Trans. Antennas Propag., vol. XXX, no. XXX }%
{Shell \MakeLowercase{\textit{et al.}}: Bare Demo of IEEEtran.cls for IEEE Journals}
\maketitle
\begin{abstract}
Mutual coupling is a dominant systematic effect in dense reflector arrays, imprinting direction-dependent and frequency-dependent structure on embedded element patterns (EEPs) and currently limiting sensitivity in precision radio measurements. Accurate modelling of these effects requires full-wave simulations of structures that are electrically large at both the array and element levels, making conventional approaches computationally prohibitive.
We present a Method-of-Moments (MoM) framework accelerated by a fast direct solver (FDS). The rotational symmetry of reflector dishes is exploited to efficiently compress self-interaction blocks of the impedance matrix. Mutual interactions are treated using a broadband multipole decomposition that remains efficient and accurate for closely spaced elements.
We demonstrate the method on arrays of tens of reflectors from the Hydrogen Epoch of Reionization Array (HERA) telescope. To scale to larger arrays, the FDS is used to construct macro-basis functions (MBFs) from a smaller representative array and embed them within a conventional MBF scheme. This allows the first computation of EEPs for the 320-element HERA core on a 128-core workstation. \textcolor{blue}{
This work has been submitted to the IEEE for possible publication.
Copyright may be transferred without notice, after which this
version may no longer be accessible.}
\end{abstract}


\section{Introduction}
Computational electromagnetics has become a key component of modern radio astronomy, with modelled instrumental EM responses embedded directly within calibration \cite{Sokolowski2017,Wijnholds2019,Borg2020} and data-processing pipelines \cite{Kern2019,Josaitis2022,Rath2025,Ohara2025}. Imperfect knowledge of the instrumental response can introduce systematic effects that can lead to false detections or biased measurements. Among these systematics, mutual coupling (MC) between antennas has emerged as a dominant effect \cite{Sutinjo2015,deLeraAcedo2018,Barry2019,Ohara2025,HERAphase2}, as it directly alters the spatial and spectral structure of the  embedded element patterns (EEPs), limiting current observations and preventing some instruments from reaching their nominal sensitivity. Consequently, accurate and efficient modelling of MC has become a central challenge in contemporary radio astronomy experiments \cite{Maaskant2006, BuiVan2018, Bolli2022, Gueuning2022,Conradie2024}.

\begin{figure*}[h]
    \centering
    \includegraphics[
        width=\textwidth,
        trim=13cm 8.0cm 11cm 6.0cm,
        clip
    ]{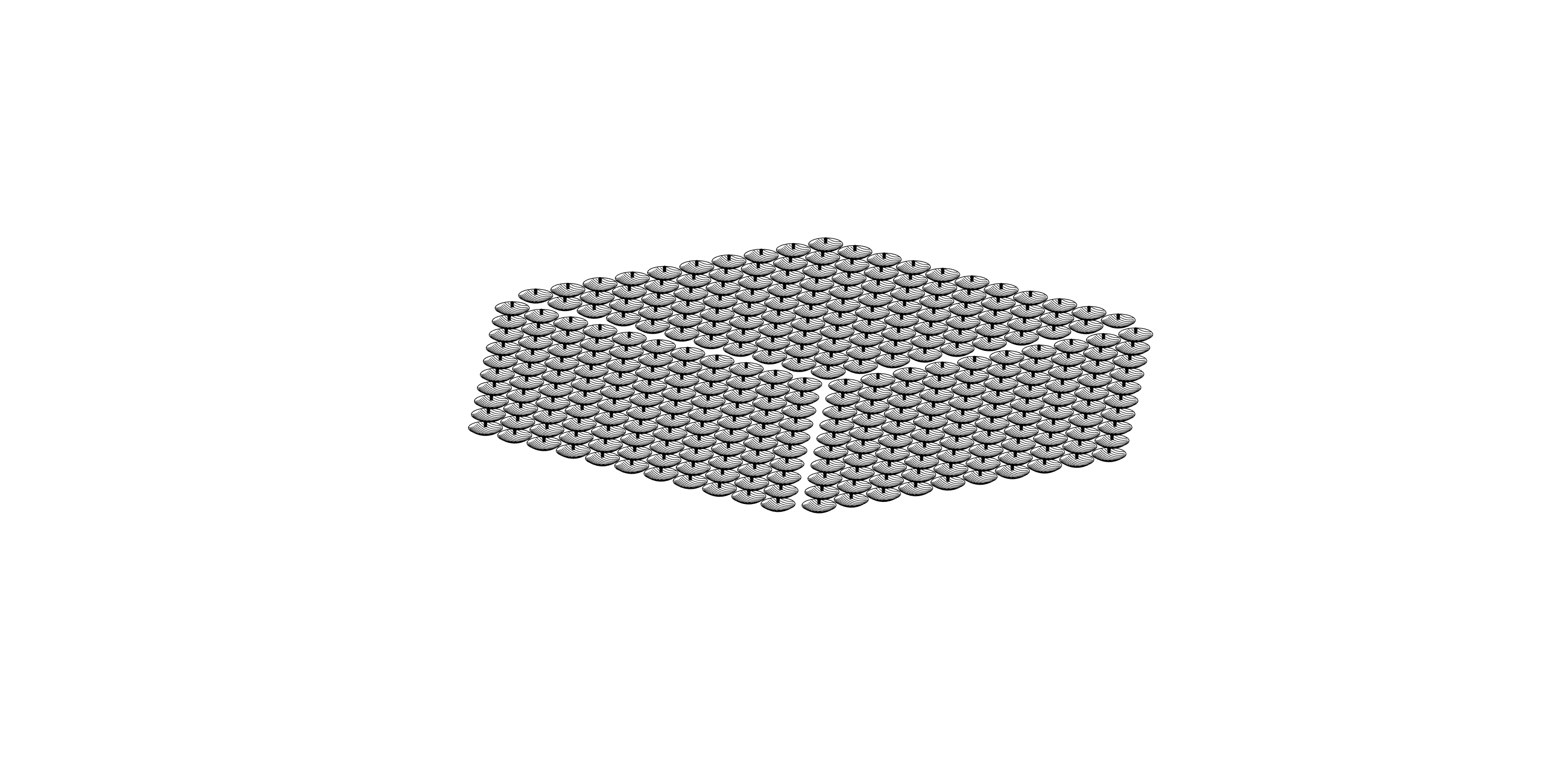}
    \caption{Visualization of the Hydrogen Epoch of Reionization Array (HERA) core. The core comprises 320 fixed, zenith-pointing 14 m parabolic reflector antennas arranged in a compact planar hexagonal configuration. Each reflector is fed by a suspended dual-polarized Vivaldi antenna. The core extends approximately 300 m in diameter and forms the densely sampled central region of the full HERA interferometric array.}
    \label{fig:HERA_core}
\end{figure*}

While EM simulation of large arrays of small antennas has been extensively studied due to its relevance in communications applications \cite{Chen2018}, closely packed arrays of reflector antennas remain largely specific to radio astronomy and comparatively underexplored. In modern dish-based radio telescopes such as CHIME \cite{Bandura2014}, Tianlai\cite{Xu2015}, HERA \cite{DeBoer2017} and HIRAX \cite{Crichton2022}, reflector diameters can span tens of wavelengths, and dishes are often arranged in compact configurations with subwavelength inter-element spacing (see an example in Fig.~\ref{fig:HERA_core}). Although asymptotic \cite{Yaghjian1981,Tian2007} and analytical \cite{Josaitis2022,Rath2025} formulations have been proposed, their validity for densely packed arrays remains uncertain, particularly with respect to accurately capturing near-field coupling, edge diffraction, and scattering arising from geometric irregularities. These limitations motivate the adoption of fast full-wave approaches, as initiated in \cite{Gueuning2022}. In this paper, we address this challenge by introducing an accelerated electromagnetic modelling technique based on the Method of Moments (MoM) for computing the EEPs of large arrays of reflector antennas. 

Reflector antennas are typically electrically large so that even an isolated element must often be discretised using many basis functions, leading to significant computational cost when using conventional MoM. A common property of reflector antennas, however, is their smooth and rotationally symmetric surface. This symmetry can be exploited using body-of-revolution methods (BoR-MoM, \cite{Mautz1979,Glisson1983,Meincke2014}) or, in the case of discrete $N$-fold rotational symmetry, a block-circulant approach \cite{Vescovo1997,Carr2002,Cavillot2020}. In the latter case, the MoM impedance matrix becomes cyclic and the problem can be decomposed into independent sector problems, avoiding construction of the full self-interaction matrix and allowing faster solutions for the individual element. In this work, we adopt this latter approach and account exactly for the presence of the feed and its strong coupling with the dish using a Schur complement. This technique is then used within the global array solve to accelerate the isolated-element computations.

Beyond this element-level acceleration, matrix filling and inversion of the global MoM system remain prohibitive for large arrays with many antennas and/or electrically large elements. Numerous acceleration strategies have been proposed for array problems, including iterative \cite{Fasenfest2004,Conradie2023,Cavillot2025} and direct solvers \cite{Zuter2000,Maaskant2008,Craeye2014FERMAT,WangChew1990,GurelChew1992,LiDingHeldringHuChenVecchi2021, GueuningTAP2022}. Iterative approaches, typically accelerated using the Multi-Level Fast Multipole Method (MLFMM), have been widely applied to single-reflector scenarios \cite{Heldring2004,Borries2014}, but may suffer from convergence issues, particularly for multiscale feed or reflector geometries. Direct approaches are attractive for array simulation, as they provide responses for all excitation ports simultaneously and avoid such convergence difficulties.
Two major families of direct solvers have been developed: fast direct solvers (FDS,\cite{LiDingHeldringHuChenVecchi2021}) and macro-basis function (MBF,\cite{Craeye2014FERMAT}) methods. Fast direct solvers emerged in the early 1990s \cite{WangChew1990} by exploiting reduced-dimensional representations of inter-element interactions, typically through modal expansions of scattered fields. A representative example is the transition-matrix (T-matrix) approach \cite{WangChew1990,GurelChew1992}, in which each cell is characterised by a mapping between incident and scattered field modes, allowing array interactions to be formulated in a reduced modal space. Introduced in the early 2000s \cite{Zuter2000}, MBF methods instead perform the reduction at the current distribution level by constructing tailored basis functions defined over one or several antenna domains. While both frameworks reduce the size of the system to be inverted, they differ in how the compression is performed: MBF methods project the MoM matrix onto a reduced current space spanned by macro-basis functions, whereas FDSs compress the exchanged fields between subdomains by exploiting the low-rank structure of mutual interactions, allowing structured factorizations and direct inversion via matrix lemmas. While MBF methods can greatly compress the MoM systems \cite{Craeye2014FERMAT}, they require a dedicated pre-processing stage to generate the macro-basis functions, whereas field-based approaches avoid this step by using generic field expansions, but at the cost of typically larger reduced systems.

In this work, we present a single-level fast direct solver (FDS) that exploits the $N$-fold rotational symmetry of the reflector surface together with the low-rank structure of mutual interactions. The self-interaction block is compressed using block-circulant operations, while mutual interactions are represented through a broadband multipole/inhomogeneous plane-wave (IPW) decomposition that preserves accuracy even for subwavelength dish–rim separations. The resulting MoM system is then efficiently inverted via a variation of the Woodbury matrix identity.
For a 33 element array, the proposed single level FDS has computational time and memory requirements comparable to the MLFMM solution obtained with Altair FEKO. However, the proposed FDS avoids the convergence issues inherent to iterative solvers when considering multi-scale geometries. In particular, FEKO's MLFMM solution does not converge for the realistic HERA Vivaldi feed model, whereas the proposed FDS remains robust. For larger arrays with hundreds of elements, the FDS is leveraged to generate a compact set of MBFs from a smaller hexagonal array. With the MBF framework, the EEPs of the 320-element HERA core are computed in approximately 5 hours per frequency point on a 128-core workstation and require 1 TB of peak memory.

The paper is organised as follows. Section~\ref{sec:AccelMoM} presents the accelerated MoM formulation. Section~\ref{sec:NumExamples} validates the approach against commercial solvers and reports numerical results for the HERA telescope. Conclusions are given in Section~\ref{sec:Conclusion}.

\section{Accelerated Method of Moments}
\label{sec:AccelMoM}

In the Method of Moments (MoM), the discretisation of the integral equations leads to a dense linear system
\begin{equation}
\mathbf{Z}\mathbf{I} = \mathbf{V},
\label{MoM}
\end{equation}
where $\mathbf{Z} \in \mathbb{C}^{N_b \times N_b}$ is the impedance matrix, $N_b$ denotes the number of basis functions, $\mathbf{I}$ contains the unknown current coefficients, and $\mathbf{V}$ is the excitation vector. In array analysis, the impedance matrix must often be solved for multiple right-hand sides, such that $\mathbf{I}$ and $\mathbf{V}$ become matrices of size $N_b \times N_{p}$, where $N_{p}$ denotes the number of array ports. We can also partition the unknowns according to individual antennas. For an array of $N_a$ reflector antennas, the global impedance matrix exhibits a block structure,
\begin{equation}
\mathbf{Z} =
\begin{bmatrix}
\mathbf{Z}_{11} & \cdots & \mathbf{Z}_{1N_a} \\
\vdots & \ddots & \vdots \\
\mathbf{Z}_{N_a1} & \cdots & \mathbf{Z}_{N_aN_a}
\end{bmatrix},
\label{globalmatrix}
\end{equation}
where $\mathbf{Z}_{ii}$ represents the self-interaction of antenna $i$, and $\mathbf{Z}_{ij}$ with $i \neq j$ describes the mutual interaction between distinct antennas.

For large array problems, both matrix assembly and inversion become computationally demanding. In fact, for electrically large reflector antennas, even assembling the matrix blocks is costly, as the number of basis functions per antenna, $N_{ba}$, can reach several tens of thousands. Two levels of acceleration are therefore introduced. Operations involving the self-interaction blocks $\mathbf{Z}_{ii}$ are accelerated by exploiting the rotational symmetry of the dish surface while the mutual interactions $\mathbf{Z}_{ij}$ are efficiently factorised using a broadband plane-wave multipole expansion. As described in the following, these complementary strategies allow rapid factorisation of the global MoM system \eqref{globalmatrix}.

\subsection{Evaluation of the self interactions}
\label{subsec:selfint}
The diagonal block $\mathbf{Z}_{ii}$ of the global impedance matrix \eqref{globalmatrix} represents the self-interaction of a single reflector antenna, comprising $N_f$ feed basis functions and $N_d$ reflector basis functions, with $N_{ba} = N_d + N_f$. As the reflector surface typically spans tens of wavelengths, we often have $N_d \gg N_f$, so both matrix assembly and local inversion are dominated by the dish contribution. We therefore partition the system into dish and feed degrees of freedom using a Schur complement (step 1), isolating the dish contribution for accelerated assembly and inversion via block-circulant operations (step 2).

\subsubsection{Dish--feed separation via Schur complement}

For a single reflector antenna, the self-interaction matrix can be partitioned as
\begin{equation}
\mathbf{Z}_{ii} =
\begin{bmatrix}
\mathbf{Z}_{\mathrm{f}} & \mathbf{Z}_{\mathrm{fd}} \\
\mathbf{Z}_{\mathrm{df}} & \mathbf{Z}_{\mathrm{d}}
\end{bmatrix},
\label{Z_reflector}
\end{equation}
where $\mathbf{Z}_{\mathrm{f}} \in \mathbb{C}^{N_f \times N_f}$ and 
$\mathbf{Z}_{\mathrm{d}} \in \mathbb{C}^{N_d \times N_d}$ denote the self-impedance matrices of the feed and dish, respectively, and 
$\mathbf{Z}_{\mathrm{fd}} \in \mathbb{C}^{N_f \times N_d}$ and $\mathbf{Z}_{\mathrm{df}} \in \mathbb{C}^{N_d \times N_f}$ describe their mutual interaction. 
The current and excitation vectors are partitioned as $\left[\mathbf{I}_{\mathrm{f}}, \mathbf{I}_{\mathrm{d}}\right]^{\mathsf{T}}$ and $\left[\mathbf{V}_{\mathrm{f}}, \mathbf{V}_{\mathrm{d}}\right]^{\mathsf{T}}$, where the subscripts f and d denote feed and dish quantities. Eliminating the dish unknowns yields a reduced feed system defined by the Schur complement of $\mathbf{Z}_{\mathrm{d}}$:
\begin{equation}
\mathbf{S}
=
\mathbf{Z}_{\mathrm{f}}
-
\mathbf{Z}_{\mathrm{fd}}
\mathbf{Z}_{\mathrm{d}}^{-1}
\mathbf{Z}_{\mathrm{df}} .
\label{eq:schur_feed}
\end{equation}
The feed current is then given by
\begin{equation}
\mathbf{I}_{\mathrm{f}}
=
\mathbf{S}^{-1}
\left(
\mathbf{V}_{\mathrm{f}}
-
\mathbf{Z}_{\mathrm{fd}}
\mathbf{Z}_{\mathrm{d}}^{-1}
\mathbf{V}_{\mathrm{d}}
\right),
\end{equation}
while the reflector currents follow as
\begin{equation}
\mathbf{I}_{\mathrm{d}}
=
\mathbf{Z}_{\mathrm{d}}^{-1}
\left(
\mathbf{V}_{\mathrm{d}}
-
\mathbf{Z}_{\mathrm{df}}
\mathbf{I}_{\mathrm{f}}
\right).
\end{equation}
These expressions provide an exact solution for the coupled feed–reflector system after eliminating the dish degrees of freedom. This methodology allows us to use distinct approaches for the feed and reflector: in our case, rotational symmetry is exploited for the dish, while the feed is treated in full generality. Note that this approach correctly models the multiple complex interactions between the feed and dish without approximation.

\subsubsection{Rapid inversion of the dish self-impedance matrix using block-circulant structure}

For a reflector surface with $N_{sec}$-fold rotational symmetry (see Fig.~\ref{fig:HERA_mesh}), discretised into $N_{sec}$ identical angular sectors each containing $N_{bsec}$ basis functions, the dish self-impedance matrix 
$\mathbf{Z}_{\mathrm{d}} \in \mathbb{C}^{(N_{sec}N_{bsec})\times(N_{sec}N_{bsec})}$
becomes block-circulant when the unknowns are ordered sector by sector,
\begin{equation}
\mathbf{Z}_{\mathrm{d}} =
\begin{bmatrix}
\mathbf{Z}_{\mathrm{d},0} & \mathbf{Z}_{\mathrm{d},1} & \cdots & \mathbf{Z}_{\mathrm{d},N_{sec}-1} \\
\mathbf{Z}_{\mathrm{d},N_{sec}-1} & \mathbf{Z}_{\mathrm{d},0} & \cdots & \mathbf{Z}_{\mathrm{d},N_{sec}-2} \\
\vdots & & \ddots & \vdots \\
\mathbf{Z}_{\mathrm{d},1} & \mathbf{Z}_{\mathrm{d},2} & \cdots & \mathbf{Z}_{\mathrm{d},0}
\end{bmatrix}.
\label{blockcirc}
\end{equation}
Hence, the matrix is fully determined by its first block row. Rather than forming $\mathbf{Z}_{\mathrm{d}}^{-1}$ explicitly, we rapidly compute its action
$\mathbf{Y}=\mathbf{Z}_{\mathrm{d}}^{-1}\mathbf{X}$ on a right-hand-side matrix
$\mathbf{X}\in\mathbb{C}^{(N_{sec}N_{bsec})\times N_p}$ partitioned by sector as
$\mathbf{X}=[\mathbf{X}_0^\top \ \cdots \ \mathbf{X}_{N_{sec}-1}^\top]^\top$,
with $\mathbf{X}_k\in\mathbb{C}^{N_{bsec}\times N_p}$.
Exploiting the block-circulant structure of $\mathbf{Z}_{\mathrm d}$,
we apply a discrete Fourier transform along the sector index
to diagonalise the system in angular modes \cite{Vescovo1997,Carr2002}.
Defining
\begin{equation}
\widehat{\mathbf{X}}_m =
\sum_{k=0}^{N_{sec}-1}
\mathbf{X}_k e^{-j2\pi mk/N_{sec}},
\end{equation}
and applying the same transformation to the sector blocks
$\mathbf{Y}_k$ and $\mathbf{Z}_{\mathrm{d},k}$ to obtain
$\widehat{\mathbf{Y}}_m$ and $\widehat{\mathbf{Z}}_{\mathrm{d},m}$,
the system decouples into $N_{sec}$ independent problems
\begin{equation}
\widehat{\mathbf{Z}}_{\mathrm{d},m}\widehat{\mathbf{Y}}_m
=
\widehat{\mathbf{X}}_m
\end{equation}
where $ m=0,\ldots,N_{sec}-1$. The sector-domain solution is recovered via the inverse transform
\begin{equation}
\mathbf{Y}_k =
\frac{1}{N_{sec}}
\sum_{m=0}^{N_{sec}-1}
\widehat{\mathbf{Y}}_m e^{j2\pi mk/N_{sec}} .
\end{equation}
The inversion of the dense $(N_{sec}N_{bsec})\times(N_{sec}N_{bsec})$
matrix $\mathbf{Z}_d$ is thus replaced by $N_{sec}$ independent systems
of size $N_{bsec}\times N_{bsec}$ together with forward and inverse
Fourier transforms along the sector index. A similar approach also allows the rapid application of $\mathbf{Z}_{\mathrm{d}}$ itself.

\subsection{Evaluation of the mutual interactions}
\label{subsec:mutint}

The mutual interactions are evaluated using a spectral factorisation of the impedance blocks $\mathbf{Z}_{ij}$, in which each interaction is expressed as a superposition of inhomogeneous plane waves (IPWs). This representation separates the antenna-dependent radiation patterns from the baseline-dependent translation operator. The use of IPWs provides numerical stability at sub-wavelength separations, as they accurately capture reactive near-field components, while requiring only a limited number of spectral samples for quasi-planar structures.

The $(m,n)$ entry of the impedance block $\mathbf{Z}_{ij}$, representing the integral reaction between antennas $i$ and $j$ associated with testing function $m$ and basis function $n$, admits the following spectral representation in terms of IPWs \cite{Jandhyala,Gueuning2025}:
\begin{align} \nonumber
[ \mathbf{Z}_{ij} ]_{mn}
=
\frac{1}{k \eta }
\iint\limits_{\Gamma}
\mathbf{f}_{t,m}(k_z,\alpha)
\cdot
\mathbf{f}_{b,n}(k_z,\alpha)\, \\
T(k_\rho P_{ij},\alpha-\phi_{ij})
\,\mathrm{d}\alpha \mathrm{d}k_z ,
\label{spectralreaction}
\end{align}
where $k$ is the free-space wavenumber and $\eta$ is the free-space wave impedance. 
The spectral representation is expressed in terms of the wavevector $\mathbf{k}$, decomposed into its vertical component $k_z$ and transverse component $k_\rho$ such that $|\mathbf{k}|=k$ and $k_\rho=\sqrt{k^2-k_z^2}$; the transverse component is parameterised by the azimuthal angle $\alpha$. 
The spectral variables, i.e. the vertical wavenumber $k_z$ and the azimuth $\alpha$, are integrated along prescribed contours $\Gamma$ in their respective complex planes \cite{Gueuning2025}. 
The quantities $\mathbf{f}_{t,m}$ and $\mathbf{f}_{b,n}$ denote the spectral radiation patterns of the testing and basis functions, respectively, while $T$ is the multipole plane-wave translation operator of the 2D Green's function \cite{Engheta}, which depends on the inter-antenna distance $P_{ij}$ in the array plane and the relative azimuth $\phi_{ij}$ between antennas $i$ and $j$.
Since the patterns are transverse to the wavevector, their product reduces to transverse electric and transverse magnetic contributions,
\begin{align}
\mathbf{f}_{t,m}\cdot \mathbf{f}_{b,n}
=
f_{tTE,m}\, f_{bTE,n}
+
f_{tTM,m}\, f_{bTM,n}.
\label{patternproduct}
\end{align}
Note that these patterns depend only on the antenna geometry and can therefore be precomputed for all basis functions prior to matrix assembly.

The spectral integral \eqref{spectralreaction} is evaluated numerically using $N_{ipw}$ IPWs. 
The TE and TM components of the testing and basis patterns of the antenna are assembled into the matrices
$\boldsymbol{\mathcal{F}}_{tTE,i}$, 
$\boldsymbol{\mathcal{F}}_{tTM,i}$,
$\boldsymbol{\mathcal{F}}_{bTE,i}$, and
$\boldsymbol{\mathcal{F}}_{bTM,i}
\in \mathbb{C}^{N_{ba}\times N_{ipw}/2}$,
where $b$ and $t$ denote basis and testing functions, respectively, and $N_{ba}$ is the number of basis functions per antenna. With these notations, the interaction block between antennas $i$ and $j$ admits the factorised representation
\begin{align}
\mathbf{Z}_{ij}
&\simeq
\boldsymbol{\mathcal{F}}_{tTE,i}\boldsymbol{\mathcal{T}}_{ij}\boldsymbol{\mathcal{F}}_{bTE,j}^{T}
+
\boldsymbol{\mathcal{F}}_{tTM,i}\boldsymbol{\mathcal{T}}_{ij}\boldsymbol{\mathcal{F}}_{bTM,j}^{T},
\label{farfacto}
\end{align}
where the plane-wave translation matrix 
$\boldsymbol{\mathcal{T}}_{ij}\in\mathbb{C}^{N_{ipw}/2\times N_{ipw}/2}$ 
is diagonal due to the separability with respect to source and observation coordinates.
For compactness, the TE and TM contributions in \eqref{farfacto} can be combined by augmenting the pattern matrices into
$\mathbf{F}_{t,i}, \mathbf{F}_{b,j} \in \mathbb{C}^{N_{ba}\times N_{ipw}}$
and defining the corresponding diagonal translation matrix 
$\mathbf{T}_{ij} \in \mathbb{C}^{N_{ipw}\times N_{ipw}}$.
Any interaction block is then written as
\begin{equation}
\mathbf{Z}_{ij}
\simeq
\mathbf{F}_{t,i}
\mathbf{T}_{ij}
\mathbf{F}_{b,j}^{T}.
\label{multipolefact}
\end{equation}
This formulation reduces the original interaction block 
$\mathbf{Z}_{ij} \in \mathbb{C}^{N_{ba}\times N_{ba}}$ 
to a diagonalised product involving matrices of width $N_{ipw}$, with 
$N_{ipw} \ll N_{ba}$. 
\subsection{Inversion of the global MoM system}
\label{subsec:invglobal}

At the global array level, the impedance matrix can now be written as a low-rank correction to the block-diagonal self-interaction matrix $\mathbf{Z}_s$,
\begin{align}
\mathbf{Z}
\simeq
\mathbf{Z}_s
+
\mathbf{F}_t
\mathbf{T}
\mathbf{F}_b^{T}.
\label{lowrankZ}
\end{align}
For an array of identical antennas, the same block appears along the diagonal of $\mathbf{Z}_s$, as well as in the block-diagonal pattern matrices $\mathbf{F}_t$ and $\mathbf{F}_b$, with blocks of size $N_{ba} \times N_{ipw}$. The global translation matrix $\mathbf{T}$ is dense, with size $N_{ipw}N_a \times N_{ipw}N_a$, and collects all pairwise translation operators between antennas.  
It is important to emphasise that \eqref{lowrankZ} is conceptual, as the full matrix $\mathbf{Z}$ is never assembled explicitly.

Using a variation of the Woodbury matrix identity \cite{Henderson1981}, the inverse of the global impedance matrix \eqref{lowrankZ} can be written as
\begin{align}
\mathbf{Z}^{-1}
\simeq
\mathbf{Z}_{s}^{-1}
-
\mathbf{Q}_t
\mathbf{T}
\mathbf{S}^{-1}
\mathbf{Q}_b^{T},
\label{lowrankZinv}
\end{align}
where the scaled Schur complement is defined as
\begin{align}
\mathbf{S}
=
\mathbf{I}
+
\mathbf{Y}_{rs}\mathbf{T}.
\end{align}
The reduced self-admittance matrix
\begin{align}
\mathbf{Y}_{rs}
=
\mathbf{F}_b^{T}
\mathbf{Z}_s^{-1}
\mathbf{F}_t
\label{reducedY}
\end{align}
represents the projection of the self-interaction admittance onto the modal subspace. 
The matrices
\begin{align}
\mathbf{Q}_t
=
\mathbf{Z}_s^{-1}\mathbf{F}_t,
\qquad
\mathbf{Q}_b
=
\mathbf{Z}_s^{-1}\mathbf{F}_b
\label{QsQt}
\end{align}
can be interpreted as macro-testing and basis function matrices obtained by solving the isolated self-interaction problem for each IPW incidence.
Multiplying \eqref{lowrankZinv} by the excitation matrix $\mathbf{V}$ yields the decomposition of the current
\begin{align}
\mathbf{I}
=
\mathbf{I}_{iso}
-
\mathbf{I}_{mut},
\label{mutcorr}
\end{align}
where the isolated contribution is $\mathbf{I}_{iso} = \mathbf{Z}_s^{-1}\mathbf{V}$ and the mutual coupling correction is $\mathbf{I}_{mut} = \mathbf{Q}_t \mathbf{I}_{r}$ with the reduced coefficient given by
\begin{align}
\mathbf{I}_{r}
=
\mathbf{T}
\mathbf{S}^{-1}
\mathbf{V}_r,
\label{Ir}
\end{align}
where the reduced excitation matrix is $\mathbf{V}_r = \mathbf{Q}_b^{T}\mathbf{V}$.

The isolated current $\mathbf{I}_{iso}$ is obtained using the block-circulant approach described in Section~\ref{subsec:selfint}. For the mutual coupling correction $\mathbf{I}_{mut}$, the pattern matrices $\mathbf{F}_{t,i}$ and $\mathbf{F}_{b,i}$ are first evaluated. The reduced self-admittance matrix $\mathbf{Y}_{rs,i} = \mathbf{F}_{b,i}^{T}\mathbf{Z}_{ii}^{-1}\mathbf{F}_{t,i}$ is then computed efficiently using the same block-circulant approach. The diagonal translation matrices $\mathbf{T}_{ij}$ are computed for each antenna pair and assembled into $\mathbf{T}$. The Schur complement $\mathbf{S}$ is formed and inverted using LU decomposition and applied to $\mathbf{V}_r$ to obtain the reduced coupling coefficients $\mathbf{I}_r$. The correction current $\mathbf{I}_{mut}$ is finally reconstructed through multiplication by $\mathbf{Q}_t$.

The computational complexity is dominated by the inversion of the scaled Schur complement, which scales as $\mathcal{O}\big((N_a N_{ipw})^3\big)$.
In the following examples, the number of plane waves $N_{ipw}$ is about two orders of magnitude smaller than the number of basis functions per antenna $N_{ba}$, leading to a substantial reduction in computational cost compared with direct inversion of the full MoM matrix~\eqref{globalmatrix}.

\section{Numerical Example: The Hydrogen Epoch of Reionisation Array (HERA)}
\label{sec:NumExamples}

\begin{figure*}[t]
    \centering
    \includegraphics[
        width=\textwidth,
        trim=15cm 8.5cm 13cm 8.0cm,
        clip
    ]{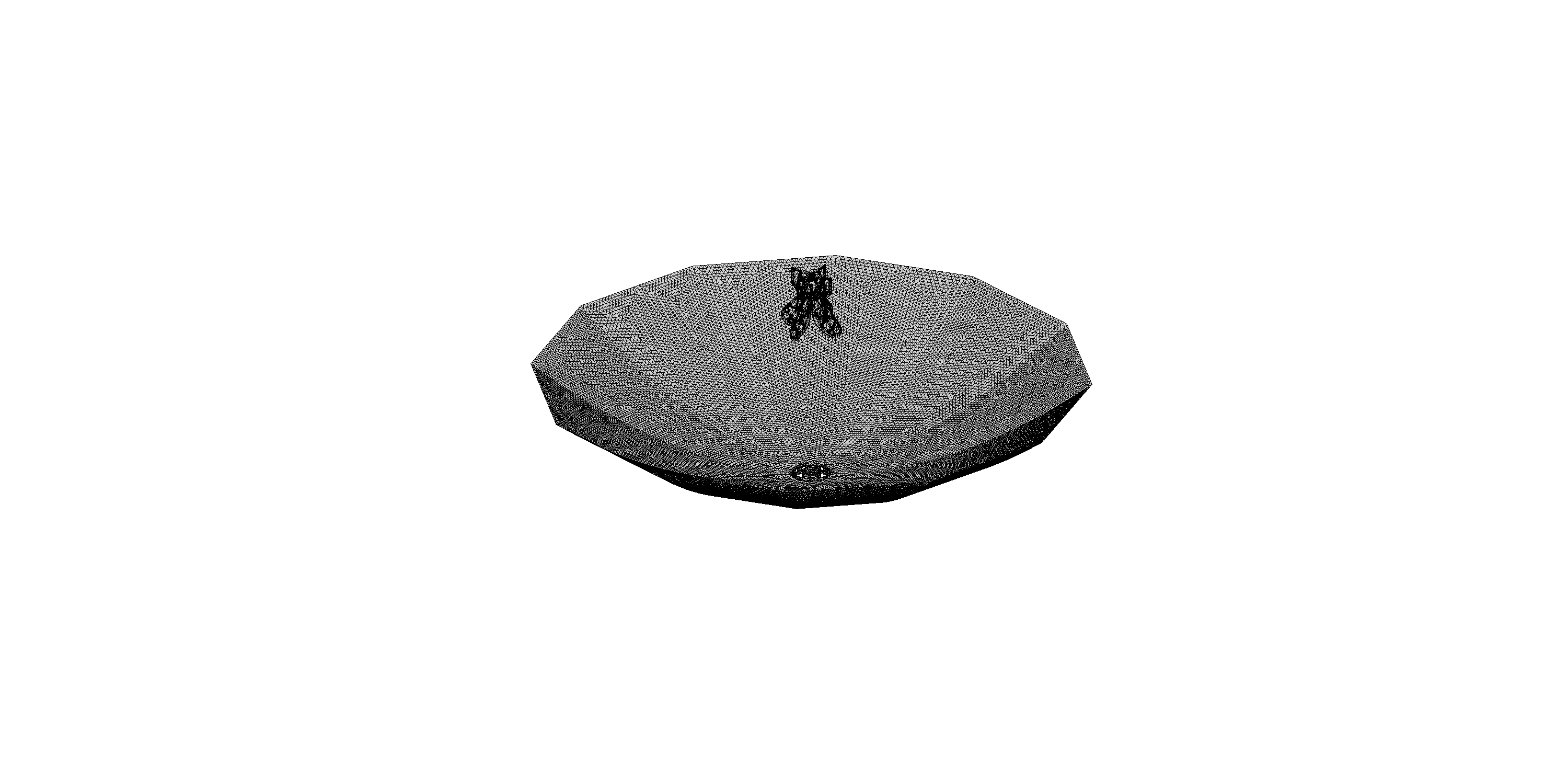}
    \caption{Surface mesh of the HERA reflector antenna used in the simulations. The model includes the 14-m diameter faceted parabolic reflector and the suspended dual-polarised Vivaldi feed located approximately 5\,m above the dish vertex. The reflector uses a focal ratio of about 0.32 and is discretised into 12 sectors with a central opening along the symmetry axis for cable routing. The dish mesh and the Vivaldi feed comprises $N_d = 62712$ and  $N_f = 43510$ RWG basis functions, respectively.}
    \label{fig:HERA_mesh}
\end{figure*}

\begin{figure}[t]
    \centering
    \includegraphics[width=\columnwidth,  trim=4cm 4.0cm 2cm 4.0cm,
        clip]{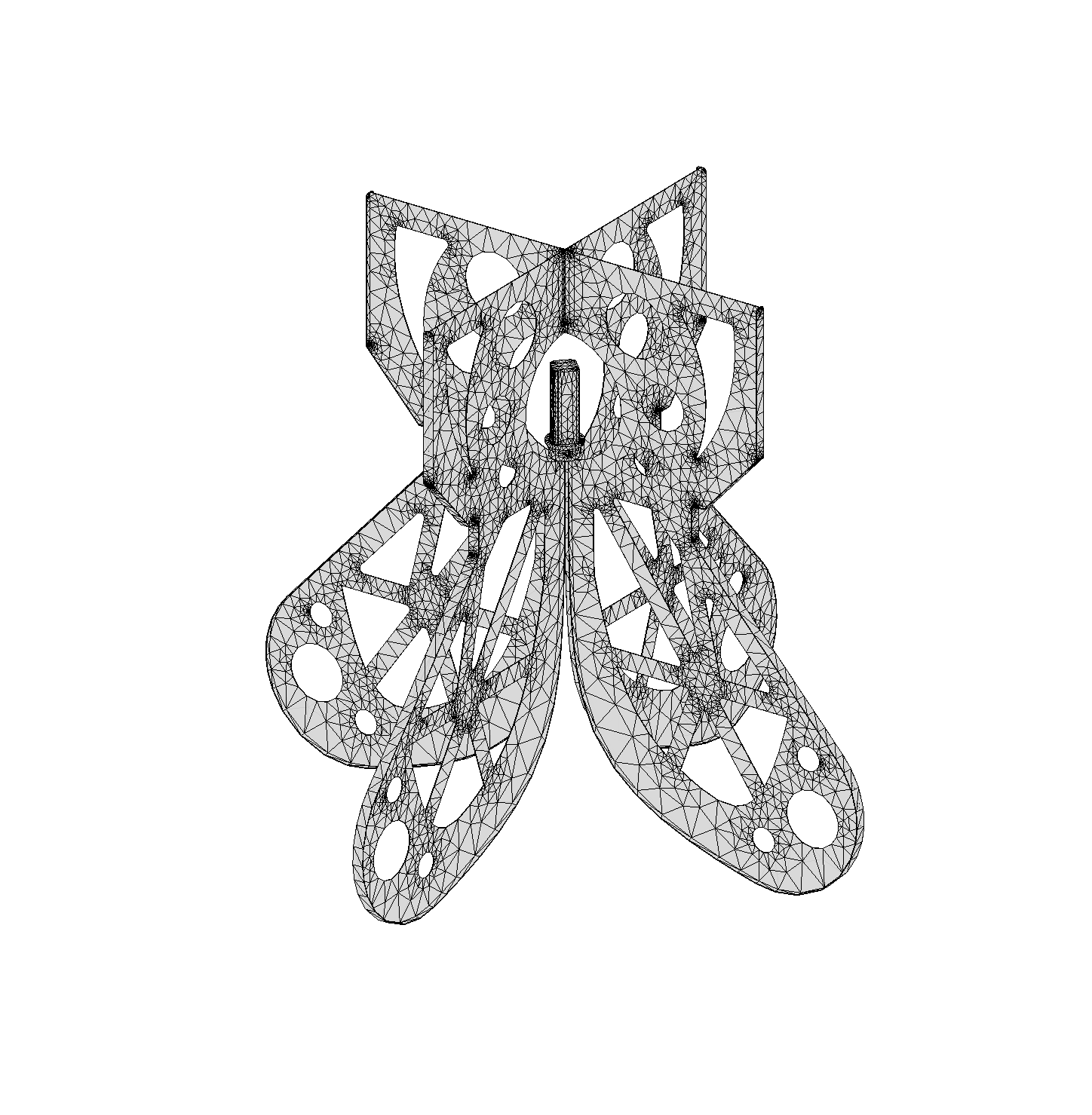}
    \caption{Surface mesh of the dual-polarised HERA feed \cite{Fagnoni}. The feed is a wideband Vivaldi antenna developed for Phase II to illuminate the 14-m reflector over  50--250 MHz.}
    \label{fig:HERA_feedmesh}
\end{figure}

The Hydrogen Epoch of Reionisation Array (HERA \cite{DeBoer2017}) is a radio interferometer designed to detect the redshifted 21-cm signal from the Epoch of Reionisation. The instrument is composed of zenith-pointing 14-m-diameter parabolic reflectors arranged in a compact hexagonal configuration with three radial branches providing access to the array (Fig.~\ref{fig:HERA_core}). Each reflector is fed by a dual-polarised Vivaldi antenna with a 1.2-m footprint \cite{Fagnoni} (Fig.~\ref{fig:HERA_mesh} and Fig.~\ref{fig:HERA_feedmesh}). The dense layout leads to non-negligible mutual coupling between suspended feeds, with additional interactions mediated by diffraction from the reflector rims.

Recent Phase II results \cite{HERAphase2} indicate that MC is a dominant systematic preventing the instrument from reaching its theoretical thermal noise floor. The published modelling analyses of MC effects relied on a far-field, single-scattering model of the coupled-element response \cite{Kern2019,Josaitis2022,Rath2025}, which reproduces several qualitative features but is reported to underestimate the observed systematic levels \cite{HERAphase2}.
This motivates the development of full-wave modelling tools capturing dense-array mutual interactions. In the following, HERA is adopted as a representative case study to assess the accuracy and scalability of the proposed approach. We progress from simplified validation cases to simulations of the densely packed 320-element core shown in Fig.~\ref{fig:HERA_core}. 

Simulations were performed on a workstation equipped with an AMD EPYC 7662 processor (128 cores, 1.5\,GHz) and 2\,TB of RAM. Computation times and memory usage are summarised in Tables~\ref{tab:computationcosts} and \ref{tab:computationcosts_iso}. Table~\ref{tab:computationcosts} summarises the computation time for the overall array solution. The pre-computation time includes frequency interpolation of the self-interaction matrices, computation of the complex pattern matrices $\mathbf{F}_i$ \eqref{multipolefact}, construction of the modal matrices $\mathbf{Q}_t$ and $\mathbf{Q}_s$ \eqref{QsQt}, and the construction of the reduced admittance matrix $\mathbf{Y}_{rs}$ \eqref{reducedY}. The matrix filling time corresponds to construction of the Schur complement $\mathbf{S}$ \eqref{eq:schur_feed}, while the solve time corresponds to the evaluation of the mutual coupling correction $\mathbf{I}_r$ \eqref{Ir}. Table~\ref{tab:computationcosts_iso} presents the self-interaction computation costs. The filling time corresponds to the evaluation of $\mathbf{Z}_f$, $\mathbf{Z}_{df}$ \eqref{Z_reflector}, and the first row of $\mathbf{Z}_d$ \eqref{blockcirc}. The compression time refers to the construction of the reduced self-admittance matrix $\mathbf{Y}_{rs}$ \eqref{reducedY}.

\begin{table*}[t]
\centering
\setlength{\tabcolsep}{3pt}
\renewcommand{\arraystretch}{1.5}

\caption{Computational cost across the numerical examples for the reference MoM, the fast direct solver (FDS), the macro-basis-function (MBF) formulation, and FEKO’s MLFMM solution, considering both a simplified feed antenna and the full HERA antenna model (Fig.~\ref{fig:HERA_mesh}). The pre-computation time of the MBF solver excludes the MBF generation phase (i.e. the FDS solve for the 19-element array). All simulations were performed on an AMD EPYC 7662 system (128 cores, 2 TB RAM).}

\begin{tabular}{ |p{3.5cm}|C|C|C|C|C|C|C| }
\hline
\textbf{Array configuration} 
& \multicolumn{2}{c|}{\textbf{2-element}} 
& \multicolumn{3}{c|}{\textbf{Hex 33}} & \textbf{Hex 19} & \textbf{HERA 320}  \\
\hline
\textbf{Feed model} 
& \multicolumn{5}{c|}{X-dipole}   & \multicolumn{2}{c|}{HERA Vivaldi} \\
\hline
\textbf{Solver} 
& ref. MoM 
& \multicolumn{2}{c|}{FDS}   
& MBF & MLFMM (FEKO) & FDS & MBF \\
\hline

\textbf{Pre-computation time (min)} 
& -  & \multicolumn{2}{c|}{1.5} & 4.5 & 5 & 30 & 19 \\

\textbf{Matrix filling time (min)} 
& 270 & 0.01 & 6 & 0.5 & -  & 1.4 & 58 \\

\textbf{Solve time (min)} 
& 54 & 0.05 & 32& 0.5 & 81.4 & 7.3 & 178 \\

\textbf{Total time (min)} 
& 324 & 1.6 & 39.5 & 5.5 & 86.4 & 38.7 & 255\\

\textbf{Peak memory (GB)} 
& 755  & 84 & 545 & 77 & 265 & 221 & 1110 \\
\hline
\end{tabular}
\label{tab:computationcosts}
\end{table*}

\begin{table}[t]
\centering
\setlength{\tabcolsep}{0.8pt}
\renewcommand{\arraystretch}{1.8}
\caption{Computation cost of the self-interaction blocks using the reference MoM (ref. MoM) and the block-circulant approach (circ. MoM). The first two columns correspond to a simplified crossed-dipole model, while the last two columns report results for the full HERA element, both meshed at 250 MHz. The reported compression time corresponds to the construction of the reduced admittance matrix $\mathbf{Y}_{rs}$ (i.e. inversion and compression of the self-interaction matrix $\mathbf{Z}_{ii}$). All simulations were performed at 150 MHz on an AMD EPYC 7662 workstation (128 cores, 2 TB RAM).}

\begin{tabular}{ |p{2.7cm}|C|C|C|C|}
\hline
\textbf{Feed model} 
& \multicolumn{2}{c|}{X-dipole } 
& \multicolumn{2}{c|}{HERA}\\
\textbf{Dish basis fns. $N_d$} 
& \multicolumn{2}{c|}{$62712$} 
&  \multicolumn{2}{c|}{$62712$} \\
\textbf{Feed basis fns. $N_f$} 
&  \multicolumn{2}{c|}{$1754$} 
&  \multicolumn{2}{c|}{$43510$}
\\ \hline
\textbf{Solver} 
& ref. MoM & circ. MoM & ref. MoM & circ. MoM  \\ \hline
\textbf{Filling time (min)} 
& 133 & 15 & 361 & 165\\
\textbf{Compress. time (min)} 
& 8.3 & 0.7 & 1.9 & 27 \\

\textbf{Peak memory (GB)} 
& 126 & 24 & 486 & 206\\
\hline
\end{tabular}
\label{tab:computationcosts_iso}
\end{table}

\begin{figure}[t]
    \centering

    \subfloat[EEP magnitude in dBV.]{
        \includegraphics[width=0.9\columnwidth]{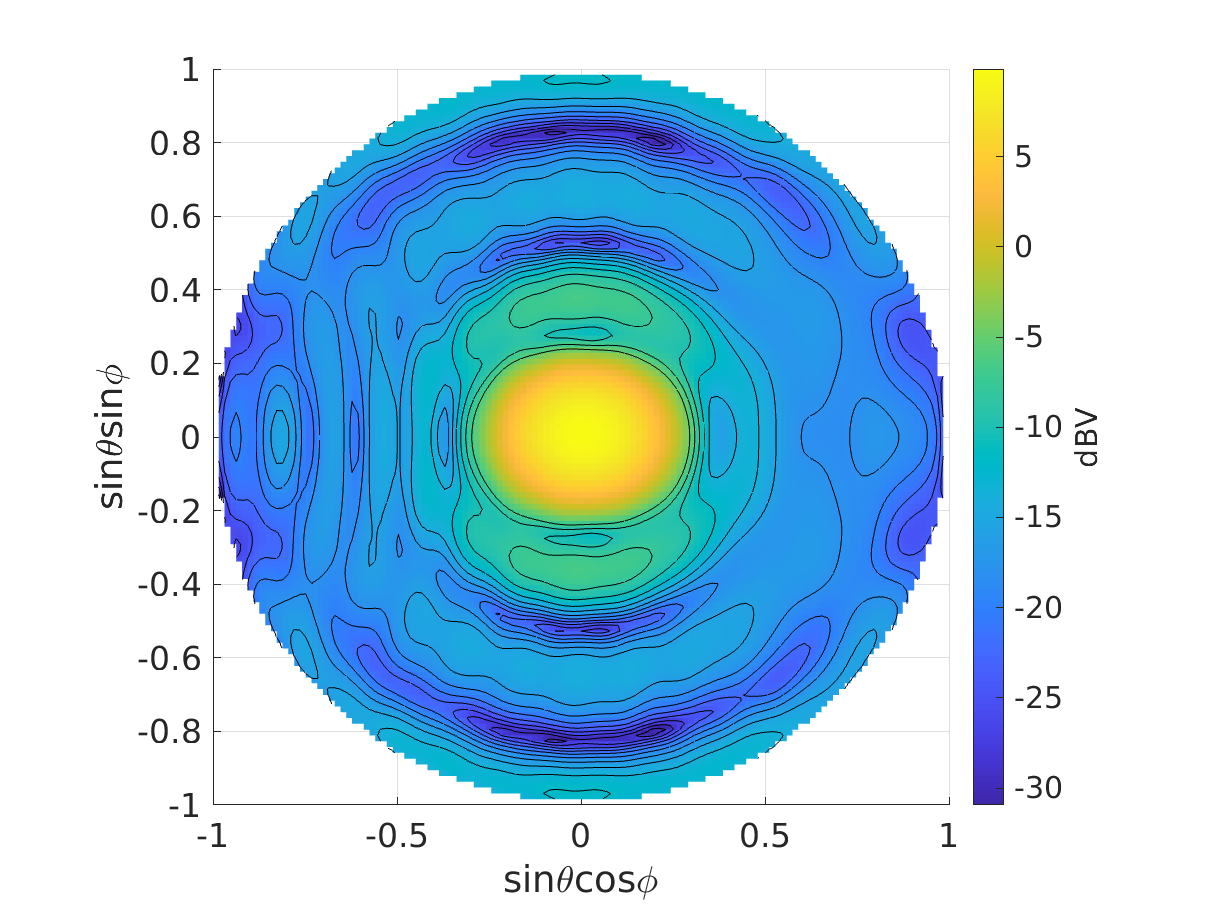}
    }
    \vspace{1mm}
    \subfloat[Residual relative to the reference MoM solution in dBV.]{
        \includegraphics[width=0.9\columnwidth]{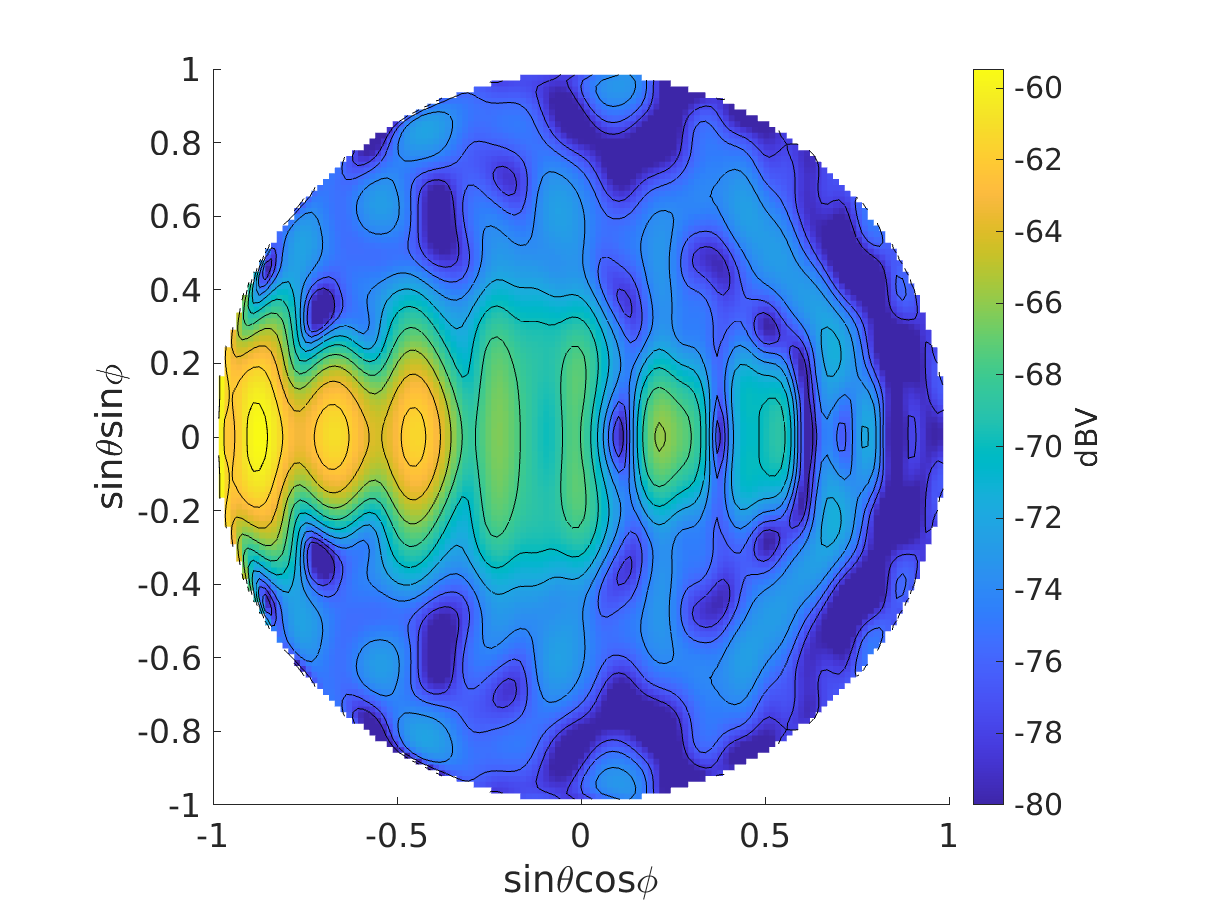}
    }
    \caption{Two-element validation at 100\,MHz. (a) EEP (ref. MoM) for the East--West feed polarisation showing background ripple on the left side of the main beam due to MC effects. (b) Residual error (FAST–ref. MoM), below $-60$\,dBV and confirming agreement to three significant digits.}
    \label{fig:hera_validation_2elem}
\end{figure}

\subsection{Validation}
Before presenting the results for the full HERA simulation, we first validate the proposed method, which incorporates the FDS module described in Section~\ref{sec:AccelMoM} and developed as part of the Fast Array Simulation Tool (FAST). Validation is first performed against a brute-force MoM reference implementation based on vectorised MATLAB routines, with low-level functions adapted from \cite{ACAsolver}. Although this implementation is not optimised for computational efficiency, it provides a reliable benchmark and was observed to be approximately $10$ times slower than FEKO’s MoM solver. The validation cases use a simple feed configuration to allow direct comparison with both the reference MoM implementation and FEKO’s MLFMM solution.  This feed consists of a 1.3\,m crossed dipole located 0.75\,m above a circular ground plane of diameter 2.25\,m. The feed resonates at 100\,MHz and is discretised using 1754 RWG basis functions. For all validation cases, the reflector considered is that of the HERA antenna (Fig.~\ref{fig:HERA_mesh}), discretised at $\lambda/10$ at 250 MHz, resulting in 62712 RWG basis functions.

Table~\ref{tab:computationcosts_iso} shows that, for the simple feed antenna, the block-circulant approach is 9 and 11 times faster than the brute-force approach for the matrix filling and compression steps, respectively, while requiring five times less memory. For the full HERA model, the speed-ups are reduced because of the larger number of feed unknowns.
The self-interaction MoM blocks, 
$\mathbf{Z}_{ff}$ $\mathbf{Z}_{df}$ and $\mathbf{Z}_{d,m}$, vary slowly with frequency and can therefore be interpolated from coarse frequency samples. In the examples considered next (10\,MHz sampling interpolated to a 1\,MHz sweep), this reduces the cost of assembling the MoM matrix by approximately a factor of 10. The construction of the reduced admittance matrix $\mathbf{Y}_{rs}$ (the compression step) must nevertheless be performed at the finer frequency sampling, since it involves the inverse MoM matrix that is not necessarily frequency-smooth.

The first validation case considers a two-element configuration with 14\,m reflectors separated by 14.6\,m centre-to-centre along the East–West axis, corresponding to a 60\,cm rim-to-rim gap (approximately $\lambda/5$ at 100\,MHz). The EEP for the East--West feed polarisation is shown in Fig.~\ref{fig:hera_validation_2elem}(a). Small ripples are visible on the left side of the main beam, arising from mutual coupling with the neighbouring antenna. The residual error, defined as the magnitude of the complex difference between the FAST and reference MoM electric fields, is shown in Fig.~\ref{fig:hera_validation_2elem}(b). The error remains below $-60$\,dBV, while the EEP sidelobes are around $-10$\,dBV, corresponding to roughly three significant digits of agreement. Note that the approximation error only arises from the factorisation of the mutual interaction blocks (Section~\ref{subsec:mutint}), as the self-interaction blocks are computed exactly with the block-circulant approach (Section~\ref{subsec:selfint}). The reference MoM simulation requires $5.5$ hours and $755$\,GB of memory, whereas FAST requires $1.6$ min and $84$ GB of RAM.

\begin{figure}[t]
    \centering
    \includegraphics[width=\columnwidth]{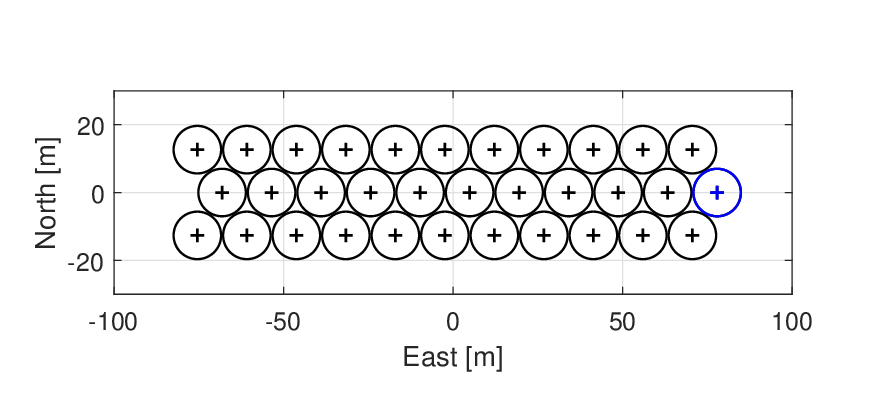}
    \caption{Hexagonal layout of a 33-element HERA reflector array (minimum spacing 14.6 m) used for validation against FEKO’s MLFMM solution. The array consists of three rows of 11 elements spanning 160 m, corresponding to the radius of the HERA core (see Fig.~\ref{fig:HERA_core}). The active element is highlighted in blue.}
    \label{fig:hex33}
\end{figure}

\begin{figure}[t]
    \centering
    \subfloat[East--West plane cut.]{
        \includegraphics[width=\columnwidth]{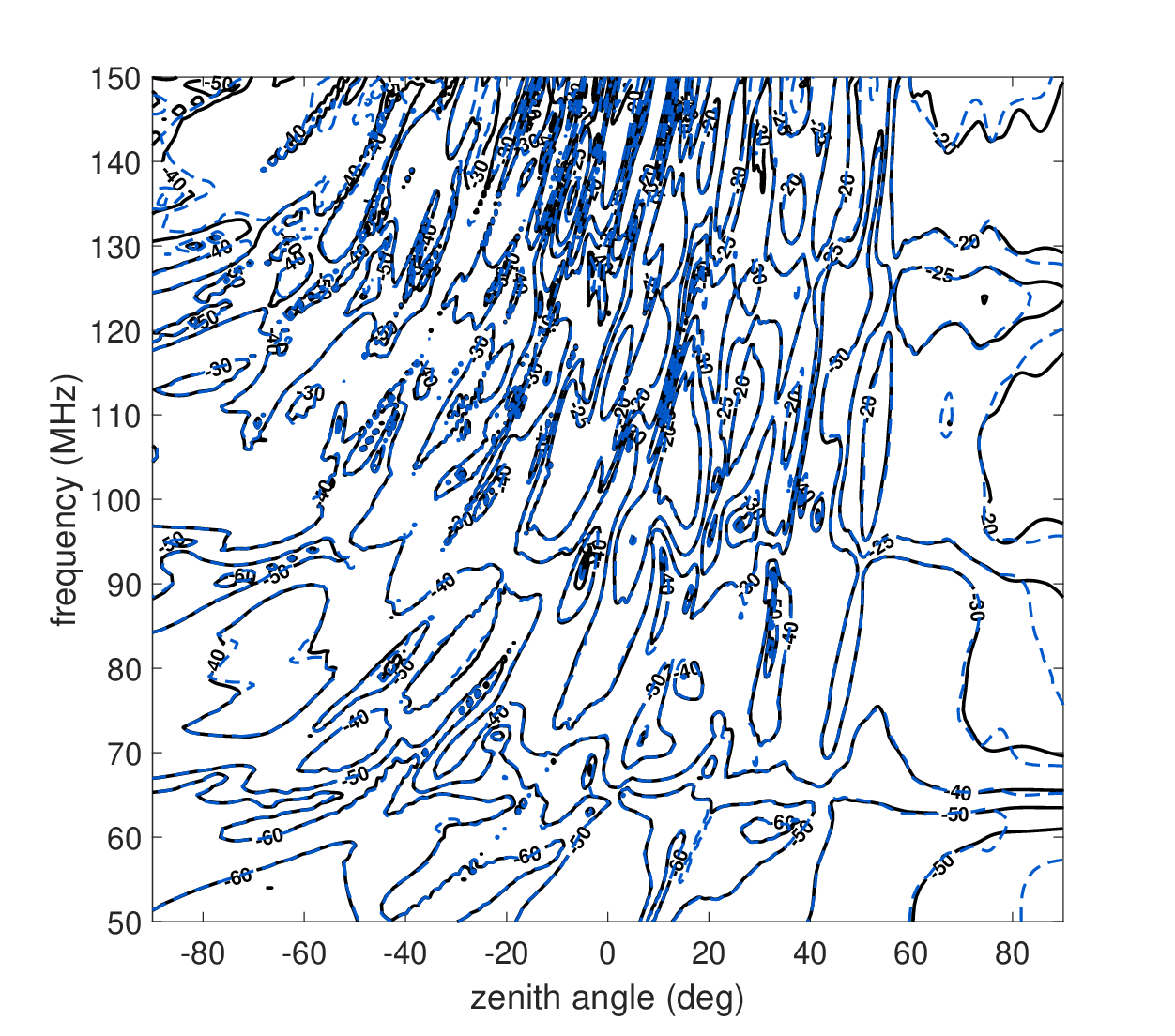}
    }
    \vspace{1mm}
    \subfloat[North--South plane cut.]{
        \includegraphics[width=\columnwidth]{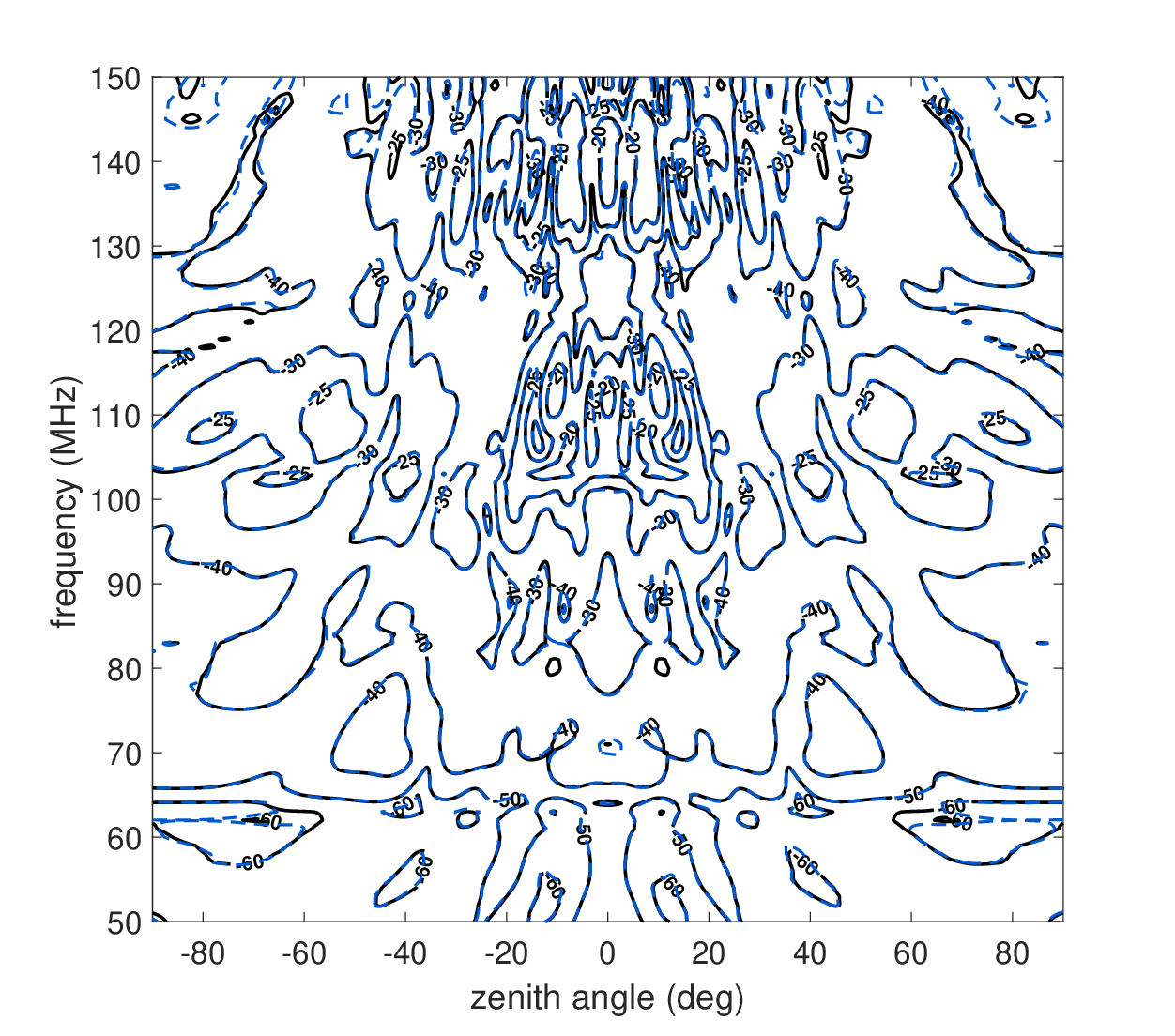}
    }
    \caption{Validation of the proposed solver against FEKO’s MLFMM solution for the embedded element pattern (EEP) of the edge element in the 33-element configuration highlighted in Fig.~\ref{fig:hex33}. The plots show waterfall maps of the residual EEP-IEP (in dBV) as a function of frequency and zenith angle for North–South feed polarisation. Black contours: proposed method (FAST). Blue dashed contours: FEKO.}
    \label{fig:hera_validation_33elem}
\end{figure}

The second validation case compares FAST with FEKO’s MLFMM EEP solution for the edge element (highlighted in blue in Fig.~\ref{fig:hex33}) of a 33-element array arranged in three rows of 11 elements, over a frequency sweep from $50$ to $150$\,MHz with 1\,MHz sampling. To isolate fine-scale MC effects, we analyse here the difference between the isolated element pattern (IEP) and the EEP. This residual corresponds to the radiation pattern associated with the MC correction $\mathbf{I}_{mut}$ \eqref{mutcorr}. Figure~\ref{fig:hera_validation_33elem} shows waterfall plots with contour overlays versus frequency and zenith angle for both East--West and North--South cuts, using the North--South polarised feed. The FEKO results (blue contours) and FAST results (black contours) show good agreement, with nearly coincident contour lines. 
For the computation of all $33\times2$ EEPs, FDS required 40 minutes compared with 86 minutes for FEKO’s MLFMM (with a residual threshold of $10^{-4}$). However, FEKO required 265\,GB of memory, whereas the FDS approach required approximately twice as much (545\,GB). Thus, for this intermediate example, the FDS is approximately twice as fast but requires roughly twice the memory. As shown in the following section, the FDS has the advantage of providing solutions for complex multi-scale geometries for which FEKO's fast iterative solver fails to converge.

\begin{figure*}[t]
    \centering
    
    \includegraphics[width=\textwidth,  trim=2.0cm 0cm 2.0cm 1cm, clip]{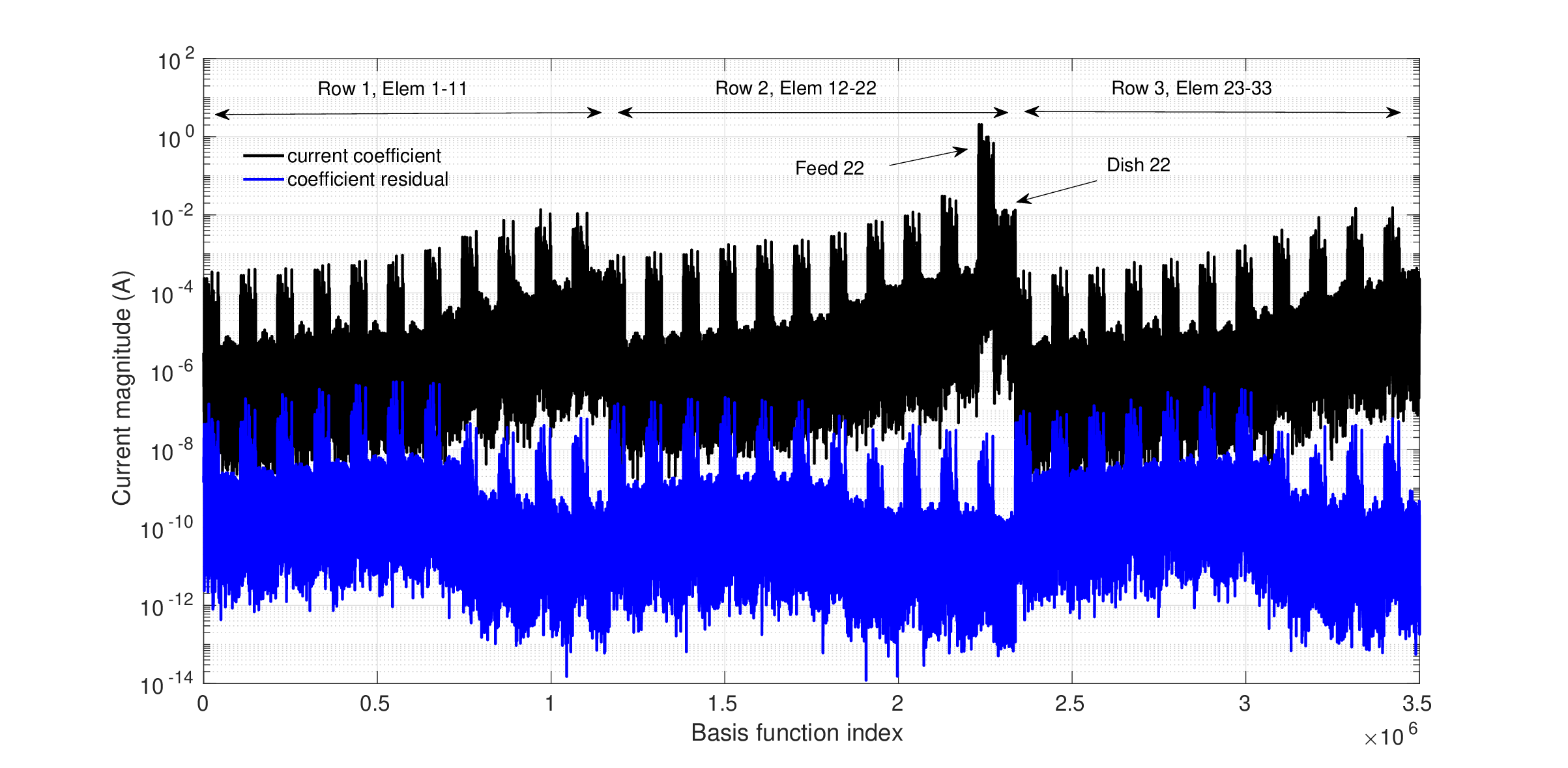}
    \caption{Validation of the MBF approach by comparison with the FDS solution for the MoM current coefficients of edge element number 22 in the 33-element layout shown in Fig.~\ref{fig:hex33} at 150\,MHz. Black: coefficient magnitudes; blue: MBF–FDS residual. Basis functions are ordered by row, with feed unknowns followed by dish unknowns for each element. The residual confirms agreement between the two solvers to three significant digits.}
    \label{fig:MBF_validation}
\end{figure*}

\begin{figure*}[t]
    \centering
    \subfloat[East--West plane cut.]{
        \includegraphics[width=\textwidth, trim=0.0cm 0 0.0cm 0, clip]{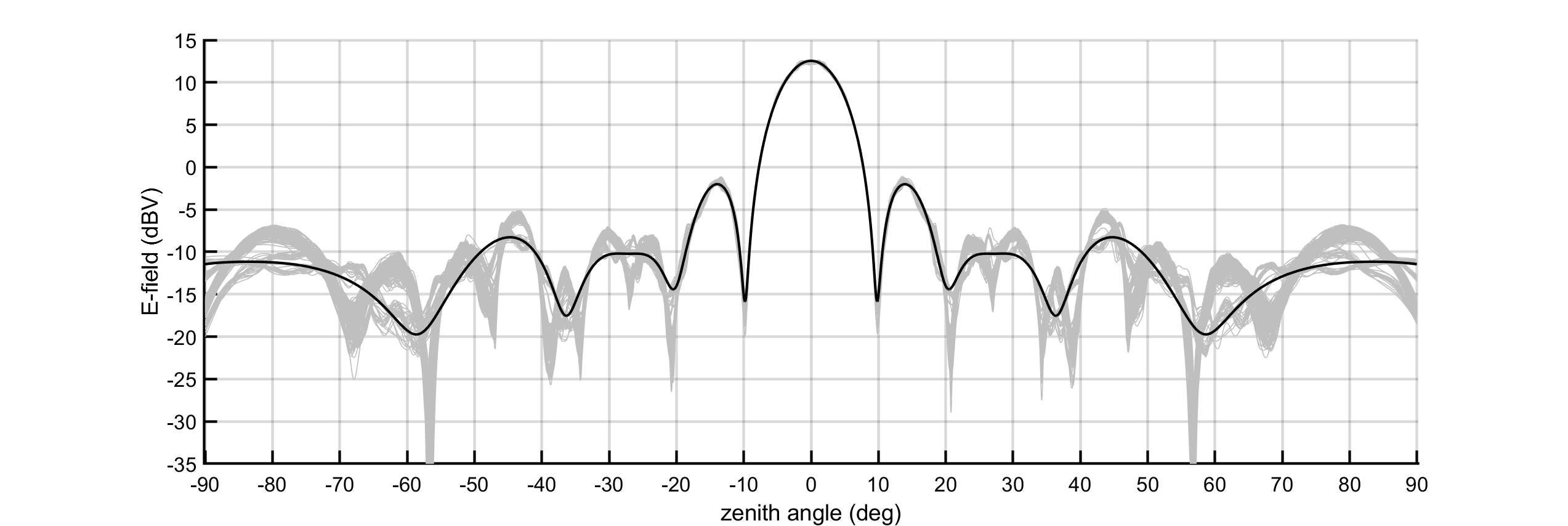}
    }%
    \\
    \subfloat[North--South plane cut.]{
        \includegraphics[width=\textwidth, trim=0.0cm 0 0.0cm 0, clip]{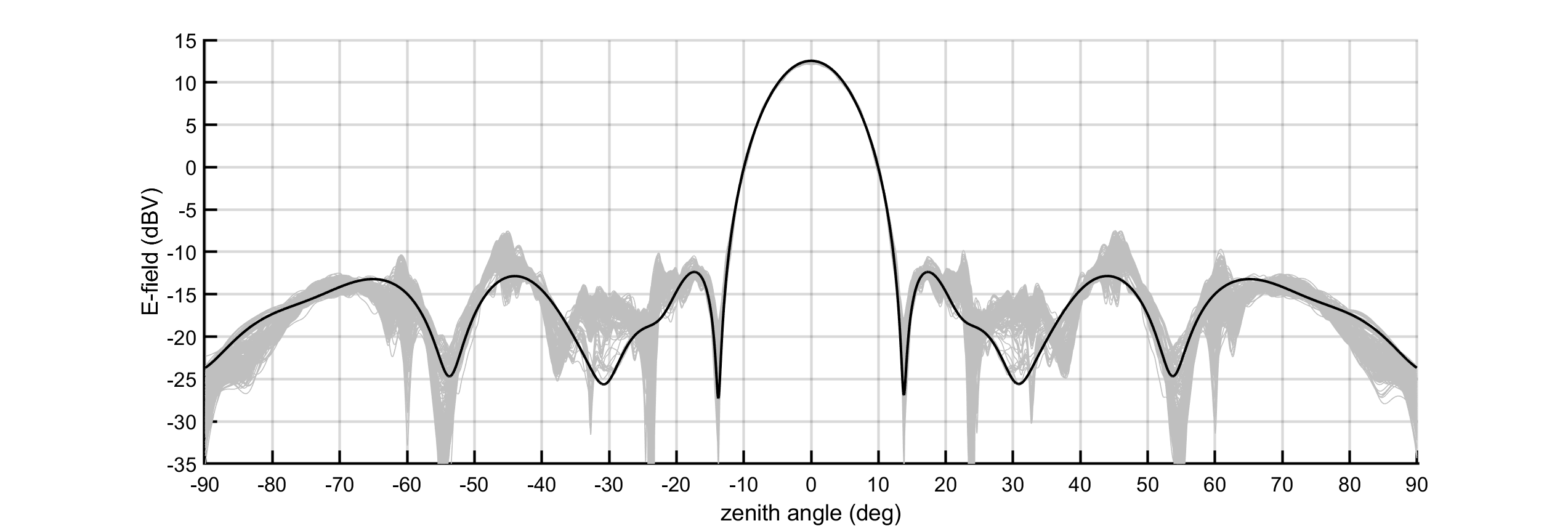}
    }
    \caption{The 320 embedded element patterns of the HERA core (see Fig.~\ref{fig:HERA_core}) at 150,MHz for North–South feed port excitation. Grey: all EEPs; black: isolated element pattern (IEP). Mutual coupling effects are primarily visible in the sidelobes, approximately -20\,dBV below the main beam value, where rapid spatial variations and a series of deep nulls that are absent from the IEP are observed.}
    \label{fig:hera_core_EEPs}
\end{figure*}

\subsection{The 320-element core of the HERA telescope}
\label{sec:HERAsims}
We now consider the HERA Vivaldi feed model, discretised with 43510 RWG basis functions (Fig.~\ref{fig:HERA_feedmesh}). Compared with the simplified feed case, for which the condition number of the MoM matrix $\mathbf{Z}_{ff}$ was $2.3\times10^3$, the full feed matrix has a much higher condition number of approximately $1.8\times10^{10}$, due to the multiscale geometry of the feed antenna. Consequently, FEKO's MLFMM iterative solver did not converge with any available preconditioner. In contrast, the FDS solver constructs and inverts the feed matrix exactly via a Schur complement and provides an accurate solution. Agreement with the reference MoM for the two element configuration is at the level of three significant digits, as observed with the simplified feed results in Fig.~\ref{fig:hera_validation_2elem}.

As shown in Table~\ref{tab:computationcosts}, the FDS remains limited to arrays of a few tens of HERA antennas. For example, the 33-element case already requires about 500 GB of memory, so inversion of the reduced matrix for the 320-element array would exceed the available 2 TB of RAM. Larger arrays are therefore treated here using a MBF approach, which provides higher compression of the MoM matrix and allows direct inversion of the reduced system. Indeed, the number of required MBFs can be made significantly smaller than the number of IPWs used in the FDS by generating them from a smaller canonical array (see~\cite{Craeye2014FERMAT}). This motivates the hybrid strategy adopted here: a reduced set of MBFs is first extracted by collecting the current distributions from a 19-element hexagonal array using the FDS and then these MBFs used to compress the impedance matrix of the full 320-element core array within the MBF framework of~\cite{Gueuning2025}, which relies on the same broadband multipole expansion as in \eqref{spectralreaction}.

Let us first validate the MBF approach at 150\,MHz. The current distributions on every antenna (active and passive) for all array port excitations (N--S and E--W) yield $2 \times 19 \times 19 = 722$ candidate current modes. As these modes can be linearly dependent, they are reduced to 600 dominant modes and orthogonalised via a SVD. The comparison of MoM element current coefficients ($\mathbf{I}$ in \eqref{MoM}) for the edge element excited in a 33-element array (Fig.~\ref{fig:hex33}) shows agreement to three significant digits with those obtained using the MBF approach (Fig.~\ref{fig:MBF_validation}).
For this example, the MBF solver requires 5.5 min per frequency point and 77\,GB of memory (excluding the MBF generation step corresponding to the full FDS solve for the 19-element array), compared with 40 min and 545\,GB for the FDS.

Finally, we analyse the solution for the full HERA core illustrated in Fig.~\ref{fig:HERA_core}. 
Figure~\ref{fig:hera_core_EEPs} shows the EEPs for all 320 elements at 150\,MHz for North--South feed excitation (light grey), plotted along the East--West and North--South cuts. The black curve denotes the isolated element pattern (IEP). While the main-beam levels appear similar across the EEPs and the IEP, clear differences arise in the sidelobe region. Mutual coupling introduces per-element variations, producing nulls in several directions and a much richer angular structure. Due to the symmetry of the array layout, some EEPs cluster into groups corresponding to elements with similar local environments (Fig.~\ref{fig:hera_core_EEPs}(a)); however, finite-array effects and the presence of the three radial branches of the layout still lead to distinct patterns for each element. MC effects typically appear about one order of magnitude (20\,dBV) below the main beam level. The computation time per frequency, including the MBF generation step, is $5$ hours using the MBF method, requiring $1.1$\,TB of peak memory to invert the reduced MBF--MoM matrix.

\section{Conclusion}
\label{sec:Conclusion}

This work presents an accelerated full-wave method for analysing mutual coupling in large irregular arrays of reflector antennas. The method is based on a Fast Direct Solver (FDS). The method incorporates two complementary acceleration strategies. First, the rotational symmetry of the reflector is exploited to compress the self-interaction blocks of the global MoM matrix efficiently. Second, all mutual interactions are factorised using a broadband inhomogeneous plane-wave decomposition. The resulting FDS provides accurate modelling of the EEPs of densely packed reflector arrays at moderate computational cost, requiring from tens of minutes to a few hours of computation time per frequency point and from tens of GB up to approximately 1 TB of memory for the examples considered in this paper.

The FDS shows similar computational cost to FEKO’s MLFMM solver for arrays comprising a few tens of antennas. However, the FDS remains stable for multi-scale problems with high-fidelity feed and reflector geometries.
In particular, FEKO's iterative solver did not converge for the realistic HERA Vivaldi model considered here due to the high conditioner number of the MoM matrix of the feed antenna, whereas the FDS produced accurate EEP solutions. For geometries where both solvers could be applied, the EEP–IEP residuals show good agreement between the two approaches. Larger arrays comprising several hundred reflectors cannot be simulated with either the MLFMM or FDS solver on the available workstation and were therefore analysed using the MBF approach presented in \cite{Gueuning2025}, with the MBF generation step accelerated using the FDS solver.

Using this hybrid MBF-FDS framework, we performed the first full-wave electromagnetic simulation of the 320-element core of the Hydrogen Epoch of Reionization Array (HERA) telescope. The results show rapid spatial variations in the EEPs, particularly in the sidelobe region, where coupling effects appear approximately 20 dBV below the main beam at 150 MHz. More generally, the proposed method provides a framework for full-wave modelling of next-generation radio interferometers composed of large reflector arrays, supporting calibration, systematic-error mitigation, and forward modelling for precision radio astronomy observations.

\appendices
\section*{Acknowledgment}
This research was supported by the Science and Technology Facilities Council (STFC). Q. Gueuning was supported by grant number ST/Y000447/1 and E. de Lera Acedo by grant number ST/V004425/1.

Portions of the manuscript text were revised using an AI-based language editing tool to improve clarity and readability. All technical content and conclusions were developed by the authors.

\begin{IEEEbiography}[{\includegraphics[width=1in,height=1.25in,clip,keepaspectratio]{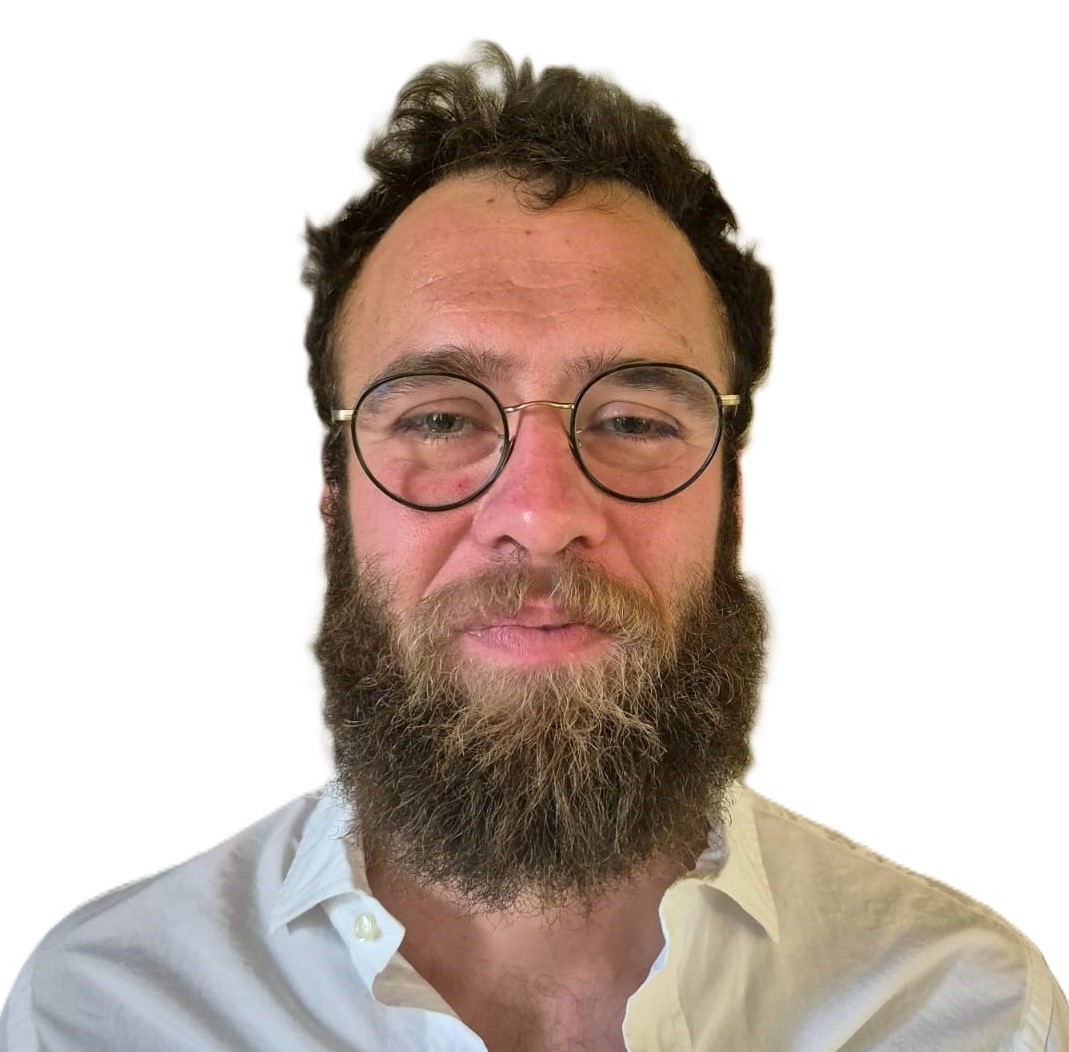}}]{Quentin Gueuning}
is an Assistant Research Professor at the University of Cambridge, UK. He is a member of the Astrophysics Group and the Kavli Institute for Cosmology at the Cavendish Laboratory and of the UK Square Kilometre Array Regional Centre (UKSRC). He has also contributed to Science Data Processor (SDP) activities for the Square Kilometre Array Observatory (SKAO). His research focuses on fast computational electromagnetics, mutual coupling in dense arrays, antenna array modelling and design, as well as calibration methods for radio astronomy instrumentation.
\end{IEEEbiography}

\begin{IEEEbiography}[{\includegraphics[width=1in,height=1.25in,clip,keepaspectratio]{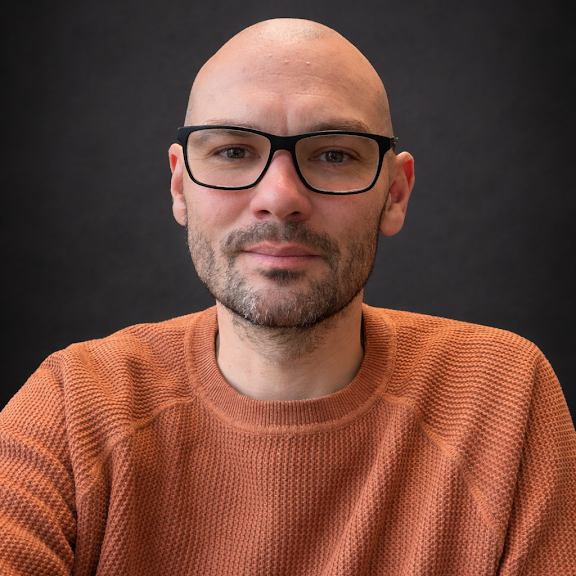}}]{Eloy de Lera Acedo} is Professor of Astrophysics at the University of Cambridge, where he leads the Cavendish Radio Astronomy and Cosmology group. His research focuses on radio instrumentation and calibration for 21-cm cosmology, and he plays a leading role in international experiments including REACH, HERA, and the SKA.
\end{IEEEbiography}

\begin{IEEEbiography}[{\includegraphics[width=1in,height=1.25in,clip,keepaspectratio]{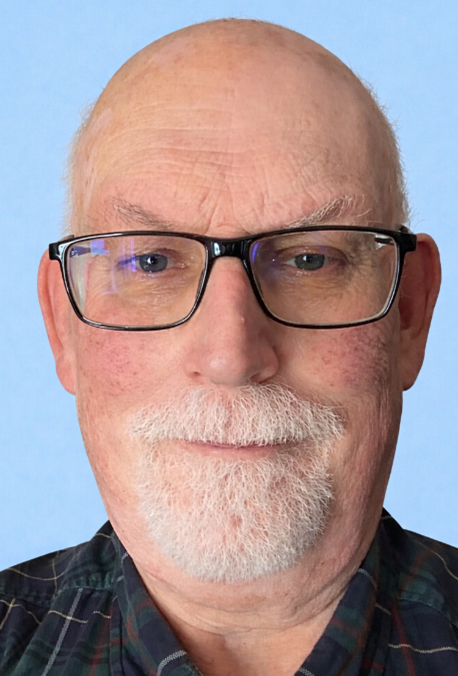}}]{Anthony K. Brown} is Emeritus Professor at Queen Mary, University of London and at the University of Manchester, UK. He is also Chief Technical Officer of Easat Radar Systems Ltd. Professor Brown is a Life Senior Member of IEEE, a Chartered Engineer and a Fellow of the Institue of Mathematics and its Applications and the IET(UK).

Tony is a former Head of the School of Electrical and Electronic Engineering at Manchester and an Asociate Dean there. He has been involved in antennas and propagation research for over 50 years in particular as applied to radar, communications and radio astronomy instrumentation. Prior to academia Tony had a long industrial career with major UK companies before forming Easat in 1987.
\end{IEEEbiography}

\begin{IEEEbiography}[{\includegraphics[width=1in,height=1.25in,clip,keepaspectratio]{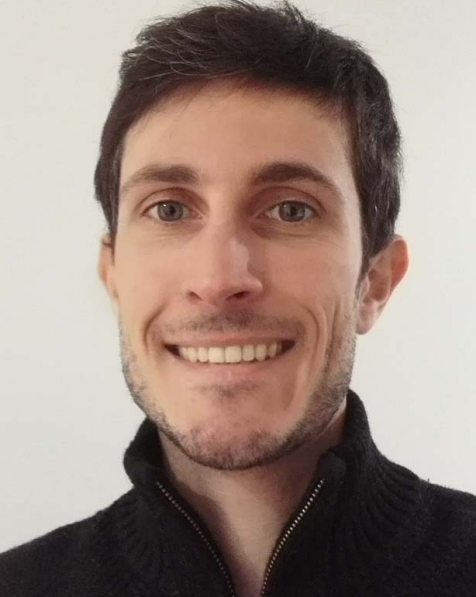}}]{Nicolas Fagnoni} received an Engineering degree in telecommunication from Telecom SudParis, as well as a Master of Science (with Distinction) in spacecraft technology and satellite telecommunication from University College London in 2011. After graduating, he worked for Eutelsat, a satellite operator, as a Satellite Service Engineer near Paris for four years. He then pursued PhD-level research in astrophysics at the University of Cambridge, supervised by Dr. Eloy de Lera Acedo. He worked on the development of new receivers for radio-telescopes, in particular for HERA and SKA, to study the conditions of the Cosmic Dawn and the Epoch of Reionization. In 2019, he joined Airbus Defence and Space UK. He is currently a Lead RF Antenna Engineer, designing antenna systems for telecom and science satellites.
\end{IEEEbiography}
\end{document}